\documentclass[prx,preprint,showpacs,preprintnumbers,amsmath,amssymb]{revtex4}

\usepackage[dvipdfmx]{graphicx}
\usepackage{bm}

\newcommand{\be}{\begin{equation}}
\newcommand{\ee}{\end{equation}}
\newcommand{\eq}[1]{Eq.~(\ref{#1})}
\newcommand{\fig}[1]{Fig.~\ref{#1}}
\def\bea{\begin{eqnarray}}
\def\eea{\end{eqnarray}}

\def\vq{{\bf q}}

\def\vk{{\bf k}}
\def\vQ{{\bf Q}}

\begin{document}

\title{Plasmarons in high-temperature cuprate superconductors} 

\author{Hiroyuki Yamase$^{1}$, Mat\'{\i}as Bejas$^{2}$, and Andr\'es Greco$^{2}$}
\affiliation{
{$^1$}International Center of Materials Nanoarchitectonics, National Institute for Materials Science, Tsukuba 305-0047, Japan\\
{$^2$}Facultad de Ciencias Exactas, Ingenier\'{\i}a y Agrimensura and Instituto de F\'{\i}sica Rosario (UNR-CONICET), Av. Pellegrini 250, 2000 Rosario, Argentina
}

\date{November 14, 2022}

\begin{abstract}
Metallic systems exhibit plasmons as elementary charge excitations. This fundamental concept was reinforced also in high-temperature cuprate superconductors  recently, although cuprates are not only layered systems but also strongly correlated electron systems. Here, we study how such ubiquitous plasmons leave their marks on the electron dispersion in cuprates. In contrast to phonons and magnetic fluctuations, plasmons do not yield a kink in the electron dispersion. Instead, we find that the optical plasmon accounts for an emergent band---plasmarons---in the one-particle excitation spectrum; acoustic-like plasmons typical to a layered system are far less effective. Because of strong electron correlations, the plasmarons are generated by bosonic fluctuations associated with the local constraint, not by the usual charge-density fluctuations. Apart from this  physical  mechanism, the plasmarons are similar to those discussed in alkali metals, Bi, graphene, monolayer transition-metal dichalcogenides, semiconductors,  diamond, two-dimensional electron systems, and SrIrO$_{3}$ films, establishing a concept of plasmarons in metallic systems in general. Plasmarons are realized below (above) the quasiparticle band in electron-doped (hole-doped) cuprates, including a region around $(\pi,0)$ and $(0,\pi)$ where the superconducting gap and the pseudogap are most enhanced. 
\end{abstract}


\maketitle
\section*{Introduction}
Superconductivity is driven by forming Cooper pairs of electrons \cite{cooper56}. This is achieved even above the boiling point of liquid nitrogen at ambient pressure in cuprate superconductors \cite{keimer15}. The mechanism of such a remarkable phenomenon has been a central issue in condensed matter physics since their discovery \cite{bednorz86}. The one-particle property of electrons may possess an important hint to solve it. In fact, electrons are not independent inside a material, but acquire the self-energy through various interactions---it is such electrons which drive the superconductivity. 

A well-known interaction is electron-phonon coupling, which renormalizes the electron band to yield a kink at the corresponding phonon energy \cite{mahan}. A kink was actually observed in not only conventional metals \cite{hengsberger99,valla99a} but also high-temperature cuprates superconductors \cite{lanzara01,zhou05}. Given that the electron-phonon coupling is the conventional mechanism of superconductivity \cite{bardeen57},  its role in  the high-$T_{\rm c}$ mechanism drew much attention. On the other hand, it is also well recognized that the formation of a kink is not a special feature of electron-phonon coupling, but rather a manifestation of coupling to bosonic fluctuations from a general point of view \cite{carbotte11}. Magnetic fluctuations are bosonic fluctuations and can in fact yield a kink in the electron dispersion at the energy of the magnetic resonance mode \cite{kaminski01,johnson01,gromko03,mou17}.

How about charge fluctuations, which are also bosonic ones? While the understanding of the charge fluctuations is crucial to the high-$T_{\rm c}$ mechanism in cuprates, it is only recently when the charge dynamics was revealed in momentum-energy space by an advanced technique of resonance inelastic x-ray scattering (RIXS). In particular,  low-energy collective charge excitations were revealed in electron-doped \cite{hepting18,lin20,hepting22} and hole-doped cuprates \cite{nag20,singh22}. The characteristic in-plane and out-of-plane  dependence allowed to identify these excitations as low-energy plasmons \cite{greco16,greco19,greco20,fidrysiak21}, which were also discussed for layered metallic systems in the 1970s \cite{grecu73,fetter74,grecu75}. Dispersing charge modes were reported previously by other groups,  too, but interpreted differently, not as plasmons \cite{ishii05,wslee14,ishii14,ishii17,dellea17}. While the dispersion was presumed to be likely an acoustic mode \cite{hepting18,nag20,lin20,singh22}, it was found very recently that the low-energy plasmons are gapped at the in-plane zone center for the infinite-layered electron-doped cuprate
${\rm Sr_{0.9}La_{0.1}CuO_{2}}$ \cite{hepting22}, in agreement with a theoretical study---the gap is predicted to be proportional to the interlayer hopping \cite{greco16}. 
These low-energy plasmons may be referred to as acoustic-like plasmons. 
While the optical plasmon itself was observed already around 1990 by electron energy-loss spectroscopy \cite{nuecker89,romberg90} and optical spectroscopy \cite{bozovic90}, these recent advances to reveal the charge excitation spectrum in cuprates attract renewed interest and plasmons offer a hot topic in the research of cuprate superconductors.

Electron-plasmon coupling was studied mainly for weakly correlated materials: 
alkali metals such as Na and Al \cite{aryasetiawan96}, Bi \cite{tediosi07}, graphene \cite{polini08,hwang08}, monolayer transition-metal dichalcogenides (Mo, W)(S, Se)$_{2}$ \cite{caruso15}, semiconductors \cite{kheifets03,caruso15,caruso15a}, and diamond \cite{caruso15a}. Here the so-called replica bands are known to be generated by coupling to plasmons; they are also referred to as plasmon satellites \cite{brar10,guzzo11,caruso15a} especially when momentum is integrated. The spectrum of the replica band is usually very broad and has low spectral weight \cite{kheifets03,markiewicz07a,hwang08,lischner13,caruso15,caruso15a,lischner15}, making it difficult to confirm it in experiments. Recently, however, the replica band was successfully resolved in various systems---graphene \cite{bostwick10,brar10,walter11}, two-dimensional electron systems \cite{dial12,jang17}, and SrIrO$_{3}$ films \cite{zliu21}. The presence of the replica band implies that the one-particle Green's function has poles. That is, the replica band corresponds to the dispersion relation of quasiparticles dubbed as plasmarons \cite{hedin67,lundqvist67,lundqvist67a,lundqvist68}. 

Can we expect plasmarons in high-temperature cuprate superconductors? This is a far from obvious issue. First of all, cuprates are strongly correlated electron systems and thus it is reasonable to distinguish cuprates from weakly correlated systems as the ones discussed in the previous paragraph---a direct analogy between them is not trivial. Second, cuprates are layered systems,  where not only the conventional optical plasmon, but also many acoustic-like plasmons are present \cite{greco16}. This is also a situation different from previous studies of plasmarons \cite{aryasetiawan96,kheifets03,tediosi07,markiewicz07a,polini08,hwang08,bostwick10,brar10,walter11,guzzo11,dial12,lischner13,caruso15,caruso15a,lischner15,jang17,zliu21}.

In this paper, we show that instead of yielding a kink, plasmons in cuprates lead to plasmarons---similar to weakly correlated systems. A common feature lies in the singularity of the long-range Coulomb interaction in the limit of long wavelength. However, the underlying physics is different. 
Instead of usual charge density-density correlations, fluctuations associated with the local constraint---non-double occupancy of electrons at any site---are responsible for the emergence of plasmarons. We find that cuprates can host plasmarons near the optical plasmon energy below (above) the quasiparticle band in electron-doped (hole-doped) cuprates. 

\section*{Results}

\subsection{Analytical scheme} 
Cuprate superconductors are doped Mott insulators---strong correlations of electrons are believed to be crucial \cite{anderson87,lee06}.  The $t$-$J$ model is a microscopic model of cuprates superconductors and is derived from the three-band \cite{fczhang88} and one-band \cite{chao77} Hubbard model. It reads
\begin{equation}
H = -\sum_{i, j,\sigma} t_{i j}\tilde{c}^\dag_{i\sigma}\tilde{c}_{j\sigma} + 
\sum_{\langle i,j \rangle} J_{ij} \left( \vec{S}_i \cdot \vec{S}_j - \frac{1}{4} n_i n_j \right)
+\frac{1}{2} \sum_{i \ne j} V_{ij} n_i n_j \,,
\label{tJV}  
\end{equation}
where $\tilde{c}^\dag_{i\sigma}$ ($\tilde{c}_{i\sigma}$) are the creation (annihilation) operators of electrons with spin $\sigma (=\uparrow, \downarrow)$  in the Fock space without double occupancy at any site---strong correlation effects,  $n_i=\sum_{\sigma} \tilde{c}^\dag_{i\sigma}\tilde{c}_{i\sigma}$ is the electron density operator, and $\vec{S}_i$ is the spin operator. While cuprates are frequently modeled on a square lattice, we take the layered structure of cuprates into account and consider a three-dimensional  lattice to describe plasmons correctly. The sites $i$ and $j$ run over such a three-dimensional lattice. The hopping $t_{i j}$ takes the value $t$ $(t')$ between the first (second) nearest-neighbor sites in the plane and is scaled by $t_z$ between the planes. The exchange interaction $J_{i j}=J$ is considered only for the nearest-neighbor sites inside the plane as denoted by $\langle i,j \rangle$---the exchange term between the planes is much smaller than $J$ (Thio {\it et al}.\cite{thio88}). $V_{ij}$ is the long-range Coulomb interaction.  

It is highly nontrivial to analyze the strong correlation effects systematically. While a variational approach is powerful \cite{spalek22}, here we employ a large-$N$ technique in a path integral representation in terms of the Hubbard operators \cite{foussats02}. In the large-$N$ scheme, the number of spin components is extended from 2 to $N$ and physical quantities are computed by counting the power of $1/N$ systematically. One of the advantages of this method is that it treats all possible charge excitations on an equal footing \cite{bejas12,bejas14}. There are two different charge fluctuations: on-site charge fluctuations describing usual charge-density-wave and plasmons, and bond-charge fluctuations describing charge-density-waves with an internal structure such as $d$-wave and $s$-wave symmetry, including the flux phase. Explicit calculations clarified that those two fluctuations are essentially decoupled to each other \cite{bejas17}. Since we are interested in plasmons, we focus on the former fluctuations. 

Because of the local constraint that double occupancy of electrons is prohibited at any lattice site, the charge fluctuations are described  by a $2 \times 2$ matrix $D_{ab}(\vq, \mathrm{i} \nu_{n})$ with $a,b=1,2$; $\vq$ is  the momentum of the charge fluctuations and $\nu_{n}$ a bosonic Matsubara frequency. While $D_{11}$ is the usual density-density correlation function, $D_{22}$ is a special feature of strong correlation effects---it describes fluctuations associated with the local constraint. As we shall clarify, this $D_{22}$ plays the central role in the formation of plasmarons. Naturally there is also the off-diagonal component $D_{12} (=D_{21})$. 

In the large-$N$ theory, $D_{ab} (\vq, \mathrm{i} \nu_{n})$ is renormalized already at leading order. After the analytical continuation $\mathrm{i} \nu_{n}\rightarrow \nu + \mathrm{i} \Gamma_{\rm ch}$, where $\Gamma_{\rm ch} (>0)$ is infinitesimally small, the full charge excitation spectrum is described by ${\rm Im}D_{ab}(\vq, \nu)$
---Bejas {\it et al}.\cite{bejas17} reported a comprehensive analysis of ${\rm Im}D_{ab}(\vq, \nu)$. In particular, ${\rm Im}D_{11}(\vq, \nu)$ predicted acoustic-like plasmon excitations with a gap at $\vq=(0,0,q_{z})$ for $q_{z}\ne 0$ as well as the well-known optical plasmon at $\vq=(0,0,0)$ \cite{greco16}. Close inspections revealed that the predicted plasmon excitations explain semiquantitatively charge excitation spectra reported by RIXS for both hole-doped  \cite{greco19,nag20} and electron-doped \cite{greco19,greco20,hepting22} cuprates. 

Charge fluctuations renormalize the one-particle property of electrons, which can be analyzed by computing the electron self-energy. This requires involved calculations in the large-$N$ theory because one needs to go beyond leading order theory. At order of $1/N$, the imaginary part of the self-energy is obtained as \cite{yamase21a} 
\be
{\rm Im}\Sigma(\vk, \omega) = \sum_{a,b=1,2} {\rm Im}\Sigma_{ab}(\vk, \omega) \,,
\label{ImSig0} 
\ee
where
\be
{\mathrm{Im}}\Sigma_{ab} ({\mathbf{k}},\omega)= \frac{-1}{N_{s} N_z}
\sum_{{\mathbf{q}}} {\rm Im}D_{ab} (\vq,\nu) h_{a}(\vk,\vq,\nu) 
h_{b}(\vk,\vq,\nu) \left[
n_{\mathrm{F}}( -\varepsilon_{\vk-\vq}) +n_{\mathrm{B}}(\nu) 
\right] \, .
\label{ImSig}
\ee
Here $\nu = \omega -\varepsilon_{\vk - \vq}$,  $\varepsilon_{\vk}$ is the electron dispersion obtained at leading order, $h_{a}(\vk, \vq, \nu)$ a vertex describing the coupling between electrons and charge excitations, $n_{\mathrm{F}}$ and $n_{\mathrm{B}}$ the Fermi and Bose distribution functions, respectively, $N_{s}$ the total number of lattice sites in each layer, and $N_{z}$ the number of layers; see Methods for the explicit forms of $D_{ab} (\vq, \nu)$, $\varepsilon_{\vk}$, and $h_{a}(\vk, \vq, \nu)$.   The real part of $\Sigma (\vk, \omega)$ is computed by the Kramers-Kronig relations. Since the electron Green's function $G(\vk,\omega)$ is written as $G^{-1}(\vk,\omega) = \omega + \mathrm{i} \Gamma_{\rm sf} -\varepsilon_{\vk} - \Sigma (\vk,\omega)$, we obtain the one-particle spectral function $A({\bf k},\omega)=-\frac{1}{\pi} {\rm Im}G({\bf k},\omega)$: 
\be
A({\bf k},\omega)= -\frac{1}{\pi} \frac{{\rm Im}\Sigma({\bf k},\omega) - \Gamma_{\rm sf}}
{[\omega- \varepsilon_{\vk}-{\rm Re}\Sigma({\bf k},\omega)]^2 
+ [{\rm Im}\Sigma(\vk,\omega) -\Gamma_{\rm sf}]^2} \,, 
\label{Akw}
\ee
where $\Gamma_{\rm sf} (>0)$ originates from the analytical continuation in the electron Green's function. 

The quasiparticle dispersion appears as poles of $A(\vk, \omega)$, i.e., a sharp peak structure of $A(\vk, \omega)$, and crosses the Fermi energy. On top of that, $A(\vk, \omega)$ can exhibit other sharp features. If plasmons themselves are responsible for yielding   additional poles of $A(\vk, \omega)$, namely fulfilling the condition of $\omega - \varepsilon_{\vk} - {\rm Re} \Sigma(\vk, \omega)=0$ with a relatively small value of $| {\rm Im} \Sigma(\vk, \omega) |$, $A(\vk, \omega)$ exhibits a peak describing electrons coupling to plasmons, namely plasmarons \cite{hedin67,lundqvist67,lundqvist67a,lundqvist68}, with a damping controlled by ${\rm Im} \Sigma(\vk, \omega)$. Note that the charge excitation spectrum also contains usual particle-hole excitations, the so-called continuum spectrum, which can also lead to poles in $A(\vk, \omega)$. Thus additional poles of $A(\vk, \omega)$ do not necessarily signal the emergence of plasmarons.  
 
While the $t$-$J$ model in \eq{tJV} contains spin fluctuations, they appear at order of $O(1/N)$ in the present theory whereas charge fluctuations appear at $O(1)$. Hence when we compute the electron self-energy at order of $1/N$, only charge fluctuations enter \eq{ImSig}, which is suitable to study the role of plasmons exclusively in the one-particle spectral function.
 
\subsection{One-particle spectral function}
A choice of model parameters is not crucial to our major conclusions. Here we present results which can be applied directly to electron-doped cuprates, especially ${\rm La_{1.825}Ce_{0.175}CuO_{4}}$ (LCCO); details of model parameters are given in Methods and
Supplementary Note 3. 

\begin{figure}[b]
\centering
\includegraphics[width=8cm]{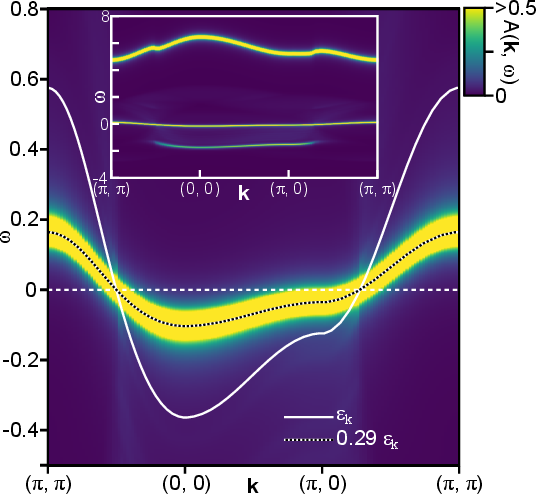}
\caption{{\bf Intensity map of the one-particle spectral function $\boldsymbol{A(\vk, \omega)}$.} The map is focused on a low-energy region along the direction $(\pi, \pi)$-$(0,0)$-$(\pi,0)$-$(\pi,\pi)$ with $k_{z}=\pi$.  For comparison, the quasiparticle dispersion obtained at leading order (solid curve) and a renormalized dispersion  multiplied by $Z=0.29$ (dotted curve) are superimposed.  The inset shows $A(\vk, \omega)$ in a wider energy window. The additional band in $\omega < 0$ corresponds to plasmarons, which we shall clarify in the present work.  
}
\label{QP}
\end{figure}

Figure~\ref{QP} shows the one-particle spectral function $A(\vk, \omega)$  along the direction $(\pi, \pi)$-$(0,0)$-$(\pi,0)$-$(\pi,\pi)$;  $k_{z}$ dependence is weak and thus $k_{z}=\pi$ is taken throughout the present paper---the value of $k_{z}$ shall be omitted for simplicity. The spectrum around $\omega=0$ is the quasiparticle dispersion renormalized by charge fluctuations. In contrast to the case of electron-phonon coupling \cite{zeyher01,zli21} and magnetic fluctuations \cite{eschrig00,markiewicz07}, it does not exhibit a kink structure. Rather it is described by the dispersion $\varepsilon_{\vk}$ (solid curve) multiplied by some constant $Z (=0.29)$ as shown by a dotted curve in \fig{QP}. This implies that  the renormalization factor depends weakly on $\vk$ and the quasiparticle spectral weight is reduced down to 0.29 by charge fluctuations.

Charge fluctuations also generate additional bands as shown in the inset in \fig{QP}. There are two major bands: a low-energy incoherent band near $\omega \approx -1.5 t$  and a high-energy side band with a large dispersion in $4 < \omega/t < 7$---the spectral weight of the former is about 10 \% and that of the latter is about 60 \%. The reason to call a side band instead of an incoherent one for the high-energy feature lies in that ${\rm Im}\Sigma(\vk, \omega)$ almost vanishes in such a high-energy region, leading to a coherent feature.  

Since the low-energy incoherent band disappears when the long-range Coulomb interaction is replaced by a short-range Coulomb interaction---the high-energy one still remains \cite{yamase21a}, a coupling to plasmons is crucial to the low-energy feature. The major point of the present work is to elucidate that the low-energy one corresponds to plasmarons \cite{hedin67,lundqvist67,lundqvist67a,lundqvist68} and is essentially the same as the so-called replica band discussed in weakly correlated electron systems \cite{aryasetiawan96,kheifets03,tediosi07,markiewicz07a,polini08,hwang08,bostwick10,brar10,walter11,guzzo11,dial12,lischner13,caruso15,caruso15a,lischner15,jang17,zliu21}.  In the following we focus on an energy window $-2 \leq \omega /t  \leq -1$. 

\subsection{Relevant contributions to the formation of plasmarons}
As seen in \eq{ImSig0},  Im$\Sigma(\vk, \omega)$ is given by the sum of four components. To elucidate the relevant contribution to forming plasmarons, we may introduce an auxiliary parameter $r (\geq 0)$ as 
\be 
{\rm Im}\Sigma(\vk, \omega; r) = {\rm Im}\Sigma_{11}+ r \times ({\rm Im}\Sigma_{22} + 2\, {\rm Im}\Sigma_{12}) \,,
\label{rSig}
\ee
where the arguments on the right hand side are omitted for simplicity and the fact that Im$\Sigma_{12}$ is equal to Im$\Sigma_{21}$ was used. The case of $r=1$ corresponds to the physical situation seen in \eq{ImSig0}. We then compute $A(\vk, \omega)$ for several choices of $r$ in \fig{rAkw}a-d, where a different color scale is used to highlight the weak feature in $r<1$; see Supplementary Note 1 for a wider energy window. Upon decreasing $r$, the incoherent  band loses intensity substantially, fades away, and finally becomes invisible in $r \lesssim 0.4$. This clearly indicates that the incoherent band is driven by components involving $a,b=2$, namely fluctuations associated with the local constraint---a direct consequence of the strong correlation effect.

\begin{figure}[tb]
\centering
\includegraphics[width=12cm]{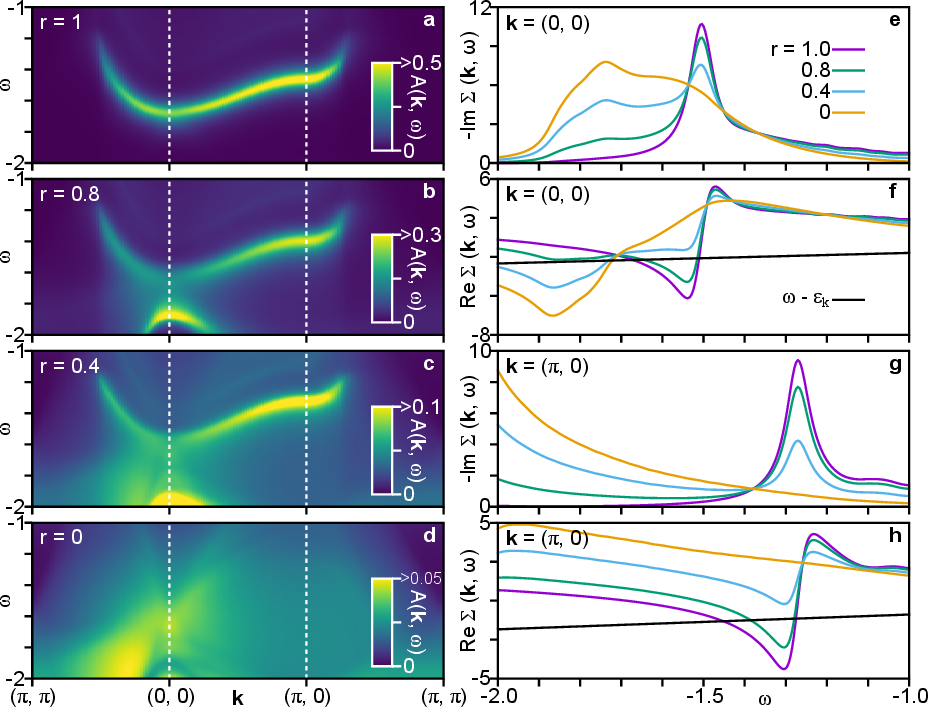}
\caption{{\bf Analysis of plasmarons in terms of \eq{rSig}.} {\bf a}-{\bf d} Intensity map of the spectral function $A(\vk, \omega)$ computed with Im$\Sigma(\vk,\omega; r)$  [\eq{rSig}] in $-2 \leq \omega/t \leq -1$ for several choices of $r$  along the direction $(\pi, \pi)$-$(0,0)$-$(\pi,0)$-$(\pi,\pi)$ with $k_{z}=\pi$. The plasmaron band in {\bf a} fades away upon decreasing $r$, indicating that fluctuations associated with the local constraint are crucially important to the plasmarons. Note a different color scale in each panel. {\bf e}-{\bf h} Imaginary and real parts of $\Sigma(\vk, \omega)$ as a function of $\omega$ at $\vk=(0, 0)$ and $(\pi,0)$ for several choices of $r$. The peak of Im$\Sigma(\vk,\omega)$ in {\bf e} and {\bf g} is determined by the optical plasmon. The line of $\omega-\varepsilon_{\vk}$ is also shown in {\bf f} and {\bf h}. The plasmaron energy is determined by its crossing point of Re$\Sigma(\vk, \omega)$ on the lower energy side when $r$ is close to 1. 
}
\label{rAkw}
\end{figure}

The corresponding ${\rm Im} \Sigma(\vk, \omega)$ and ${\rm Re} \Sigma(\vk, \omega)$ are also shown in \fig{rAkw}e-h for two choices of $\vk$. ${\rm Im} \Sigma(\vk, \omega)$ exhibits a sharp peak at $\omega \approx -1.5 t$ and $-1.3t$ for $\vk=(0, 0)$ and $(\pi,0)$, respectively, and the peak is suppressed and broadened with decreasing $r$. The peak structure of ${\rm Im}\Sigma(\vk, \omega)$ yields a large dip structure in ${\rm Re}\Sigma(\vk, \omega)$ at slightly lower energy than the peak energy of ${\rm Im}\Sigma(\vk, \omega)$ via the Kramers-Kronig relations. Consequently, the term $\omega - \varepsilon_{\vk}-{\rm Re}\Sigma(\vk, \omega)$ in \eq{Akw} can vanish at two energies when $r$ is close to 1: one is very close to the peak energy of Im$\Sigma(\vk, \omega)$ and the other corresponds to the tail of the dip structure of Re$\Sigma(\vk, \omega)$. Since Im$\Sigma(\vk, \omega)$ becomes small at the latter energy, $A(\vk, \omega)$ forms a peak there with a damping controlled by Im$\Sigma(\vk, \omega)$. Hence this peak is  incoherent, but is a resonance in the sense that $\omega - \varepsilon_{\vk} -{\rm Re}\Sigma(\vk, \omega)=0$ is fulfilled.

Which one is more crucial to the incoherent band, Im$\Sigma_{12}$ or Im$\Sigma_{22}$? To answer this, we have studied each component of Im$\Sigma_{ab}(\vk, \omega)$ and computed the spectral function for each of them. We can check that Im$\Sigma_{22}$  is responsible for the formation of the incoherent band. The component of Im$\Sigma_{12}$ works to sharpen the incoherent band by reducing the absolute value of the imaginary part of the self-energy (see Supplementary Note 2 for details).

\subsection{Role of plasmons}
What is then the role of plasmons? Plasmons are described in terms of the charge-charge correlation function, namely poles of ${\rm Im}D_{11}$ in the present theory. Because of the matrix structure of ${\rm Im}D_{ab}$, the poles are determined by the zeros of its determinant. Thus all components of ${\rm Im}D_{ab}$ contain the same poles as those in ${\rm Im}D_{11}$ and thus describe the same plasmons equally (see Figs.~1 and 8 in 
Bejas {\it et al}.\cite{bejas17} for explicit calculations). Because of the layered  structure of cuprates, plasmons have various branches depending on the  value of $q_{z}$: the usual optical plasmon corresponds to $q_{z}=0$ and the acoustic-like branches to $q_{z}\ne 0$ \cite{greco16}. Their energy varies in $0.07 \leq \nu /t \leq 1.15$ around $\vq=(0,0,q_{z})$ for the present parameters. The peak of ${\rm Im}\Sigma(\vk, \omega)$ at $\omega=-1.5 t$ ($-1.3t$) in Fig.~\ref{rAkw}e [Fig.~\ref{rAkw}g] is determined by the optical  plasmon at $\nu=1.15t$---the energy difference is easily read off from \eq{ImSig}: the energy of ${\rm Im} D_{ab}$ is given by $\nu=\omega - \varepsilon_{\vk-\vq}$, the optical plasmon is realized at $\vq={\bf 0}$, Im$D_{ab}$ is odd with respect to $\nu$, and thus the peak of Im$\Sigma(\vk, \omega)$ is shifted by $\varepsilon_{\vk-\vq}$ with $\vq={\bf 0}$.

\begin{figure}[ht]
\centering
\includegraphics[width=8cm]{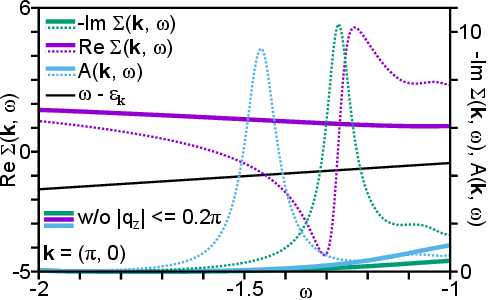}
\caption{{\bf Role of the optical plasmon for plasmarons.} The self-energy and the spectral function are computed at $\vk=(\pi, 0)$ by removing a region $| q_{z}| \leq 2\pi/10$ in the $q_{z}$ summation in \eq{ImSig}, namely without contributions from the optical plasmon. They do not form any structure. Dotted curves are the corresponding results with the full $q_{z}$ summation. The line of $\omega-\varepsilon_{\vk}$ is also given. The contrast between the solid and dotted curves demonstrates the importance of the optical plasmon to forming the plasmarons. 
}
\label{noqz0}
\end{figure}

The crucial role of the optical plasmon is also confirmed numerically. In \fig{noqz0}, we compute the self-energy Im$\Sigma(\vk, \omega)$ by removing a region $|q_{z}| \leq 2\pi/10$ in the $q_{z}$ summation in \eq{ImSig} so that the contribution from the optical plasmon becomes zero. We then observe that the peak structure in Im$\Sigma(\vk, \omega)$ completely disappears, not shifts to another energy window. The resulting $A(\vk, \omega)$ no longer forms any structure there.

It might be puzzling---on one hand, the optical plasmon is responsible for the formation of the plasmarons (\fig{noqz0}), but on the other hand, the same plasmons in ${\rm Im}D_{11}$ and ${\rm Im}D_{12}$ do not generate them (\fig{rAkw}).  The vertex function $h_{a}(\vk, \vq, \nu)$ in \eq{ImSig} was checked not to be important. The key lies in the role of the long-range Coulomb interaction $V_{\vq}$, which diverges as $q^{-2}$ in the limit of $\vq\rightarrow {\bf 0}$. As is well known, this is the very reason why the plasmons are realized \cite{mahan}---the inverse of the charge response function or the determinant of $D_{ab}$ vanishes along the plasmon dispersion. The crucial difference among ${\rm Im}\Sigma_{ab}$ appears in the numerator. To see this we study the explicit form of $D_{ab}$, which is given by 
\be
D_{ab}(\vq, \nu)=\frac{1}{\mathcal D} \left( 
\begin{array}{cc}
-\Pi_{22}(\vq, \nu) & \Pi_{12}(\vq, \nu)  - \frac{N\delta}{2} \\
 \Pi_{12}(\vq, \nu)  - \frac{N\delta}{2} \quad  &  -\Pi_{11}(\vq, \nu)  + \frac{N\delta^{2}}{2}\left( V_{\vq} -J_{\vq} \right) 
 \end{array}
\right) \,.
\ee
Here $\mathcal{D}$ is the determinant of the matrix $[D_{ab}(\vq, \nu)]^{-1}$, $\delta$ the doping rate, $J_{\vq}$ the superexchange interaction in momentum space, $\Pi_{ab}$ a bubble describing particle-hole excitations with appropriate vertex functions $h_{a}(\vk, \vq, \nu)$, and $N$ the number of spin components ($N=2$ corresponds to the physical situations); a complete expression of each quantity is given in Methods.  In the limit of $\vq \rightarrow {\bf 0}$, we can obtain by virtue of $V_{\vq} \sim q^{-2}$ 
\be
{\rm Im} D_{22}(\vq, \nu) \sim \frac{V_{\vq}^{2}  {\rm Im}\Pi_{22}}{({\rm Re}\mathcal{D})^{2} + ({\rm Im}\mathcal{D})^{2}}\,. 
\label{ImDq0}
\ee
The other components of ${\rm Im} D_{11}$ and ${\rm Im} D_{12}$ become smaller by order of $V_{{\vq}}^{-2}$ and $V_{{\vq}}^{-1}$, respectively. Hence, ${\rm Im} D_{22}$ becomes dominant over the other components in the limit of $\vq \rightarrow {\bf 0}$ and thus ${\rm Im}\Sigma_{22}$ has a sizable contribution compared with the other components. This explains the reason why ${\rm Im} \Sigma_{22}$ is responsible for the plasmarons, although all components of ${\rm Im} D_{ab}$ equally describe the same plasmons.

\section*{Discussions} 
From Eqs.~(\ref{ImSig}) and (\ref{ImDq0}), one may recognize that the mathematical structure of ${\rm Im}\Sigma_{22}$ in the limit of $\vq \rightarrow {\bf 0}$ is the same as the well-known expression of the self-energy in weak coupling theory, 
\be
{\rm Im} \Sigma^{\rm RPA} (\vk, \omega) =\frac{-1}{N_{z}N_{s}} \sum_{\vq} {\rm Im} D^{\rm RPA}(\vq, \nu) \left[
n_{\mathrm{F}}( -\varepsilon_{\vk-\vq}^{\rm RPA}) +n_{\mathrm{B}}(\nu) \right] \,,
\label{RPA}
\ee
where ${\rm Im} D^{{\rm RPA}}(\vq, \nu) = V_{\vq}^{2} {\rm Im} \Pi^{{\rm RPA}}(\vq, \nu)$ is the imaginary part of the screened Coulomb interaction computed in the random phase approximation (RPA); $\nu=\omega - \varepsilon_{\vk-\vq}^{\rm RPA}$.  Therefore weakly correlated electron systems in general can also host plasmarons in principle \cite{aryasetiawan96,kheifets03,tediosi07,markiewicz07a,polini08,hwang08,bostwick10,brar10,walter11,guzzo11,dial12,lischner13,caruso15,caruso15a,lischner15,jang17,zliu21}. However, plasmarons are  overdamped in many cases and leave faint spectral weight \cite{kheifets03,markiewicz07a,hwang08,lischner13,caruso15,caruso15a,lischner15}. This unfavorable situation is soften when the system has a relatively small band width so that the correlation effect becomes relatively large. In fact, the importance of the small band width to plasmarons was discussed in SrIrO$_{3}$ \cite{zliu21}. See Supplementary Note 5 for explicit results. 

In the present $t$-$J$ model, the band width is very small at order of $t\delta/2$ and $\delta \approx 0.1$-$0.2$.  Furthermore, the magnitude of all the components of ${\rm Im}\Sigma_{ab}$ is comparable  to each other, but the sign of ${\rm Im} \Sigma_{12}$ is the opposite to those of ${\rm Im}\Sigma_{11}$ and ${\rm Im}\Sigma_{22}$ in $\omega<0$ (see Supplementary Note 2). Hence after the summation in \eq{ImSig0}, ${\rm Im}\Sigma$ is substantially reduced in $\omega<0$. Nonetheless, a small band width allows to fulfill the resonance condition $\omega - \varepsilon_{\vk} - {\rm Re} \Sigma(\vk, \omega)=0$ as shown in Figs.~\ref{rAkw}f and h for $r=1$. These features work constructively to host plasmarons in a strongly correlated electron system  more than a weakly correlated one. 

\begin{figure}[tb]
\centering
\includegraphics[width=8cm]{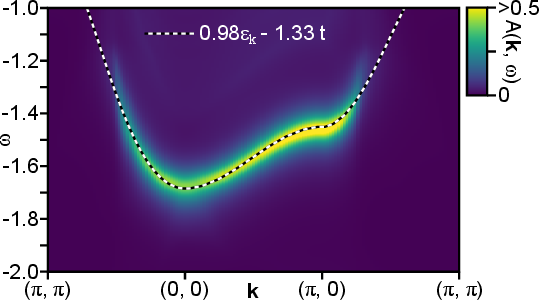}
\caption{{\bf Dispersion of plasmarons.} It follows $0.98 \varepsilon_{\vk} - 1.33t$ (dashed curve) and in this sense, it is a replica band of $\varepsilon_{\vk}$. 
}
\label{replica}
\end{figure}

The dispersion of plasmarons exhibits a dispersive feature similar to the quasiparticle dispersion as shown in \fig{replica}---the plasmaron dispersion follows $0.98 \varepsilon_{\vk} - 1.33t$. The value of $-1.33t$ is related to, but not exactly equal to, the optical plasmon energy $\nu=1.15t$ and the factor $0.98$ is a renormalization. The plasmaron dispersion can also be fitted to $\varepsilon_{\vk} -1.33t$ approximately. This feature is easily understood intuitively. The energy of the one-particle excitation $\omega$ is related to the charge fluctuation energy $\nu$ via $\omega=\nu+\varepsilon_{\vk-\vq}$ in \eq{ImSig}. Since it is the optical plasmon which generates the plasmarons, $\nu$ is estimated by its energy and $\vq$ may be put to zero; recall that Im$D_{ab}(\vq, \nu)$ is an odd function with respect to $\nu$. Consequently, the dispersion of plasmarons essentially follows the (bare) quasiparticle dispersion $\varepsilon_{\vk}$. In this sense a term of ``replica band'' used in weakly correlated electron systems can be inherited even in strongly correlated electron systems.

Figures~\ref{QP}, \ref{rAkw}a, and \ref{replica} can be applied directly to electron-doped cuprates, especially LCCO.  The energy of plasmarons is controlled by the optical plasmon energy, which can be determined precisely by electron energy-loss spectroscopy and optical spectroscopy.  Given that the typical energy scale of the optical plasmon in cuprates is around 1 eV, the plasmarons can be tested by angle-resolved photoemission spectroscopy (ARPES)  by searching the energy region typically around 1 eV below the electron dispersion especially along the direction $(0,0)$-$(\pi,0)$ [see the inset of \fig{QP} and \fig{replica}]. This energy window has not been studied in detail in ARPES \cite{armitage10}. Recalling that the plasmarons were detected even in weakly correlated systems such as graphene \cite{bostwick10,brar10,walter11}, two-dimensional electron systems \cite{dial12,jang17}, and SrIrO$_{3}$ films \cite{zliu21}, there seems a good chance to reveal them also in cuprates. 

In experiments \cite{uchida91},  the optical plasmon energy increases with carrier doping up to 20 \% doping. Therefore, the energy of plasmarons follows the same tendency. This feature can also be utilized to confirm plasmarons in cuprates. While ARPES is an ideal tool to test plasmarons, x-ray photoemission spectroscopy \cite{guzzo11} and tunneling spectroscopy \cite{brar10} can also be exploited to detect plasmarons as an emergent satellite peak. 

What happens for hole-doped cuprates, where the optical plasmon \cite{nuecker89,romberg90,bozovic90} as well as acoustic-like plasmons \cite{nag20,singh22,hepting22} typical to layered materials were actually observed? Performing the same analysis as that for the electron-doped cuprates, we predict plasmarons also in hole-doped cuprates. In contrast to the electron-doped case, however, they are realized along the direction $(\pi,0)$-$(\pi, \pi)$-$(\pi/2, \pi/2)$ in $\omega > 0$, requiring inverse ARPES to test the plasmaron dispersion; See Supplementary Note 4 for details. 

For cuprates, it has been discussed that coupling to bosonic fluctuations yields a kink in the electron dispersion. While plasmons are also bosonic fluctuations, their role in cuprates should be sharply distinguished from phonons \cite{lanzara01,zhou05} and magnetic fluctuations \cite{kaminski01,johnson01,gromko03,mou17}. Plasmons do not yield a kink (see \fig{QP}), but instead generate plasmarons as an emergent incoherent band (\fig{replica}).  

The present calculations have been performed in a layered $t$-$J$ model. If one employs a two-dimensional model, the plasmon dispersion in cuprates cannot be captured especially for the optical plasmon. In this sense, the inclusion of the three-dimensionality of the long-range Coulomb interaction is crucially important to discuss plasmarons in cuprates, although we checked that the interlayer hopping integral $t_{z}$ is not relevant to plasmarons. 

A replica band is also discussed in the polar electron-phonon coupling mechanism in TiO$_{2}$ \cite{moser13,verdi17,caruso18} and the interplay between the electron-phonon and electron-plasmon couplings was studied \cite{jalabert89}. A clear distinction between those two couplings is made by studying the carrier density dependence of the replica band \cite{zliu21}. In the present study, however, the electron-phonon coupling is irrelevant because phonon energy is limited below 100 meV in cuprates whereas our relevant energy scale is about 1 eV. 

\section*{Conclusions}
The present large-$N$ theory captures the plasmon excitations observed in both electron- and hole-doped high-temperature cuprate superconductors with a good accuracy so that detailed comparisons with experimental data were made \cite{greco19,nag20,greco20,hepting22}. We have computed the electron self-energy in the same theoretical framework, but by going beyond leading order theory. 

Our major point lies in the indication that cuprates can host plasmarons---quasiparticles coupling to plasmons---near the optical plasmon energy below (above) the quasiparticle dispersion in electron-doped (hole-doped) cuprates; plasmons do not yield a kink in the quasiparticle dispersion, in stark contrast to phonons and magnetic fluctuations. Since plasmarons are found clearly close to momentum $(\pi, 0)$, where the superconducting gap as well as the pseudogap is enhanced, it is very interesting  to explore further the role of plasmarons in the formation of the superconducting gap and the pseudogap in cuprate superconductors. 

Our second major point lies in elucidating the mechanism of plasmarons: they are driven by the strong correlation effect---fluctuations associated with the local constraint that imposes no double occupancy of electrons at any site. The underlying physics to generate plasmarons in cuprates is thus different from that in weakly correlated electron systems \cite{aryasetiawan96,kheifets03,tediosi07,markiewicz07a,polini08,hwang08,bostwick10,brar10,walter11,guzzo11,dial12,lischner13,caruso15,caruso15a,lischner15,jang17,zliu21}. However, both have a common mathematical structure to yield plasmarons, establishing a general concept of plasmarons in metals. Plasmarons tend to be well-defined for a system with a smaller band width. This condition is usually fulfilled in cuprates because of strong correlations, but also in a weakly correlated system such as SrIrO$_{3}$ films \cite{zliu21}. 

\section*{Methods} 
We present a minimal description of the large-$N$ theory of the layered $t$-$J$ model with the long-range Coulomb interaction---a complete formalism is given in
Yamase {\it et al}.\cite{yamase21a}. 

The electron dispersion $\varepsilon_{\vk}$ consists of the in-plane dispersion $\varepsilon_{\vk}^{\parallel}$ and the out-of-plane dispersion $\varepsilon_{\vk}^{\perp}$,  
\be
\varepsilon_{\vk} = \varepsilon_{\vk}^{\parallel}  + \varepsilon_{\vk}^{\perp} \,. 
\label{xik}
\ee
At leading order they are calculated as 
\be
\varepsilon_{\vk}^{\parallel} = -2 \left( t \frac{\delta}{2}+\Delta \right) (\cos k_{x}+\cos k_{y})-
4t' \frac{\delta}{2} \cos k_{x} \cos k_{y} - \mu \,,\\
\label{Epara}
\ee
\be
\varepsilon_{\vk}^{\perp} = - 2 t_{z} \frac{\delta}{2} (\cos k_x-\cos k_y)^2 \cos k_{z}  \,, 
\label{Eperp}
\ee
where $\Delta$ is the mean value of the bond field, $\delta$ the doping rate, and $\mu$ the chemical potential. For a given $\delta$, 
$\Delta$ and $\mu$ are determined self-consistently by solving the following coupled equations:  
\bea
&&\Delta = \frac{J}{4N_s N_z} \sum_{\vk} (\cos k_x + \cos k_y) n_{\mathrm{F}}(\varepsilon_\vk) \,, 
\label{Delta} \\
&&(1-\delta)=\frac{2}{N_s N_z} \sum_{\vk} n_{\mathrm{F}}(\varepsilon_\vk)\,. 
\eea

As already mentioned in the Analytical Scheme subsection, charge fluctuations in the $t$-$J$ model are composed of on-site charge and bond-charge fluctuations. They are however essentially decoupled to each other \cite{bejas17}. Since the former is relevant to the present work, we focus on that.  In this case, the bosonic propagator of charge fluctuations, namely $D_{ab}(\vq,\mathrm{i}\nu_n)$,  is described by a $2 \times 2$ matrix with $a,b=1,2$: 
\be
[D_{ab}(\vq,\mathrm{i}\nu_n)]^{-1} 
= [D^{(0)}_{ab}(\vq,\mathrm{i}\nu_n)]^{-1} - \Pi_{ab}(\vq,\mathrm{i}\nu_n)\,, 
\label{dyson}
\ee
where $D^{(0)}_{ab}(\vq,\mathrm{i}\nu_n)$ is the bare bosonic propagator, 
\begin{eqnarray}
[D^{(0)}_{ab}({\bf q},\mathrm{i}\nu_{n})]^{-1}= N \left(
 \begin{array}{cc}
\frac{\delta^2}{2} \left( V_{\vq}-J_{\vq}\right) & \frac{\delta}{2} \\
\frac{\delta}{2}  & 0 
\end{array}
\right) \,. 
\label{D0}
\end{eqnarray}
Here $J_{\vq} = \frac{J}{2} (\cos q_x +  \cos q_y)$ is the superexchange interaction in momentum space and 
$V_{\vq}$ is the long-range Coulomb interaction for a layered system \cite{becca96}: 
\be
V_{\vq}=\frac{V_{\rm c}}{A(q_x,q_y) - \cos q_z} \,,
\label{LRC}
\ee
where $V_{\rm c} = e^2 d(2 \epsilon_{\perp} a^2)^{-1}$ and $A(q_x,q_y)=\alpha (2 - \cos q_x - \cos q_y)+1$; $e$ is the electric charge of electrons, $a$ the unit length of the square lattice, $d$ the distance between the layers, $\alpha$ describes the anisotropy between the in-plane and out-of-plane interaction and is given by $\alpha=\frac{\tilde{\epsilon}}{(a/d)^2}$ with $\tilde{\epsilon}=\epsilon_\parallel/\epsilon_\perp$, where $\epsilon_\parallel$ and $\epsilon_\perp$ are the 
dielectric constants parallel and perpendicular to the planes, respectively. The $2 \times 2$ matrix $\Pi_{ab}$ is the bosonic self-energy at leading order 
\begin{eqnarray}
&& \Pi_{ab}(\vq,\mathrm{i}\nu_n)
            = -\frac{N}{N_s N_z}\sum_{\vk} h_a(\vk,\vq,\varepsilon_\vk-\varepsilon_{\vk-\vq}) 
            \frac{n_{\mathrm{F}}(\varepsilon_{\vk-\vq})-n_{\mathrm{F}}(\varepsilon_\vk)}
                                  {\mathrm{i}\nu_n-\varepsilon_\vk+\varepsilon_{\vk-\vq}} 
            h_b(\vk,\vq,\varepsilon_\vk-\varepsilon_{\vk-\vq}) \nonumber \\
&& \hspace{25mm} - \delta_{a\,1} \delta_{b\,1} \frac{N}{N_s N_z}
                                       \sum_\vk \frac{\varepsilon_\vk-\varepsilon_{\vk-\vq}}{2}n_{\mathrm{F}}(\varepsilon_\vk) \; , 
\label{Pi}
\end{eqnarray}
and the $2$-component vertex is given by  
\be
 h_a(\vk,\vq,\nu) = \left(
                    \frac{2\varepsilon_{\vk-\vq}+\nu+2\mu}{2}+
                   2\Delta \left[ \cos\left(k_x-\frac{q_x}{2}\right)\cos\left(\frac{q_x}{2}\right) +
                                  \cos\left(k_y-\frac{q_y}{2}\right)\cos\left(\frac{q_y}{2}\right) \right], 1 \right) \, .
\label{vertex-h}
\ee

We then compute the electron self-energy from charge fluctuations described by \eq{dyson} at order of $1/N$. This yields \eq{ImSig} in the main text---its derivation is elaborated in
Yamase {\it et al}.\cite{yamase21a}.

Fixing temperature to zero, we choose parameters $J/t=0.3$, $t'/t=0.3$, $t_{z}/t=0.03$, $\alpha=2.9$, $V_{\rm c}/t=18$, $\delta=0.175$, $N_{z}=10$, $\Gamma_{\rm ch}/t=0.03$, $\Gamma_{\rm sf}/t=0.03$, and $t/2=0.5$ eV, which reproduce semiquantitatively the plasmon excitations observed in RIXS for one of the typical electron-doped cuprates LCCO \cite{hepting22}. The factor of 1/2 in $t/2$ comes from a large-$N$ formalism where $t$ is scaled by $1/N$. We assume $N = 2$ in comparison with experiments. These parameters were obtained to achieve the best fit to the experimental data under an additional conditions that they should be realistic and do not contradict with the existing knowledge. While $\Gamma_{\rm ch}$ and $\Gamma_{\rm sf}$ are positive infinitesimals from the analytical point of view,  we employ small, but finite values in actual numerical calculations. This may mimics broadening of the spectrum due to electron correlations at higher orders as well as instrumental resolution. In the figures we  presented, all quantities with the dimension of energy are measured in units of $t$.

\acknowledgments
The authors thank N. P. Armitage, M. Hepting, and A. M. Ole\'s for valuable discussions. A part of the results presented in this work was obtained by using the facilities of the CCT-Rosario Computational Center, member of the High Performance Computing National System (SNCAD, MincyT-Argentina). A.G. is indebted to warm hospitality of Max-Planck-Institute for Solid State Research. H.Y. was supported by JSPS KAKENHI Grant No.~JP20H01856. 

\bibliography{main} 

\newpage


\setcounter{figure}{0}
\renewcommand{\thefigure}{S\arabic{figure}}%
\setcounter{equation}{0}
\renewcommand{\theequation}{S\arabic{equation}}%
\setcounter{subsection}{0}

\section*{Supplementary Note 1: $\boldsymbol{A(\vk, \omega)}$ in full energy window} 
\begin{figure}[ht]
\centering
\includegraphics[width=8cm]{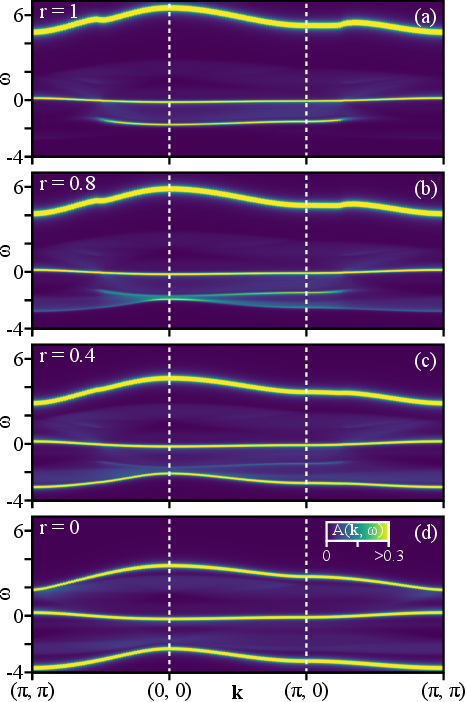}
\caption{{\bf  Intensity maps of the spectral function $\boldsymbol{A(\vk, \omega)}$ computed with Im$\boldsymbol{\Sigma(\vk,\omega; r)}$  [\eq{rSig}] in a wider energy region than Fig.~\ref{rAkw}a-d.}   
$A(\vk, \omega)$ is shown for several choices of $r$  along the direction $(\pi, \pi)$-$(0,0)$-$(\pi,0)$-$(\pi,\pi)$ with $k_{z}=\pi$. The spectral weight close to $\omega=0$ is the quasiparticle dispersion. 
}
\label{rAkw-SI}
\end{figure}

In Fig.~\ref{rAkw}, we focused on the energy window of plasmarons. For completeness, we present results in a full energy window in \fig{rAkw-SI}. The quasiparticle dispersion is always present around the zero energy, independent of $r$ as it should be. At $r=0$, where only the component $\Sigma_{11}$ contributes to the self-energy in \eq{rSig}, there are two side  bands in the positive and negative energy region. Upon introducing the effect of $\Sigma_{22}$, i.e., increasing $r$, the side band in the negative energy is completely diminished in $r \gtrsim 0.8$ and is replaced by the plasmarons. On the other hand, the side band in high-energy region ($2 <  \omega/t < 7$) is mainly controlled by $\Sigma_{11}(\vk, \omega)$ and the effect of $\Sigma_{22}(\vq, \omega)$ pushes the side band to higher energy.

\section*{Supplementary Note 2: Component analysis of $\boldsymbol{{\rm Im}\Sigma_{ab}(\vk, \omega)}$} 
\subsection{Basic property} 
As clarified in Ref.~\onlinecite{bejas17}, Im$D_{ab}(\vq, \nu)$ is an odd function with respect to $\nu$; the diagonal parts Im$D_{aa}(\vq, \nu)$ are positive for $\nu > 0$ as expected, but the off-diagonal part Im$D_{12}(\vq, \nu)$ is negative for $\nu>0$. As seen in \eq{vertex-h}, $h_{1} \propto \nu$ for a large $\nu$ and $h_{2}=1$. Therefore \eq{ImSig} implies that the off-diagonal parts of Im$\Sigma_{ab}(\vk, \omega)$ become negative (positive) for $\omega>0 (<0)$ whereas the diagonal parts are always negative as they should be. In this sense, the off-diagonal component alone does not have any physical meaning. Its role is to suppress (enhance)  the diagonal components of ${\rm Im} \Sigma_{aa}(\vk, \omega)$ in the negative (positive) $\omega$ region when all components $a$ and $b$ are summed up in \eq{ImSig0}. 

\subsection{Spectral function ${A_{ab}(\vk, \omega)}$}
\begin{figure}[th]
\centering
\includegraphics[width=8cm]{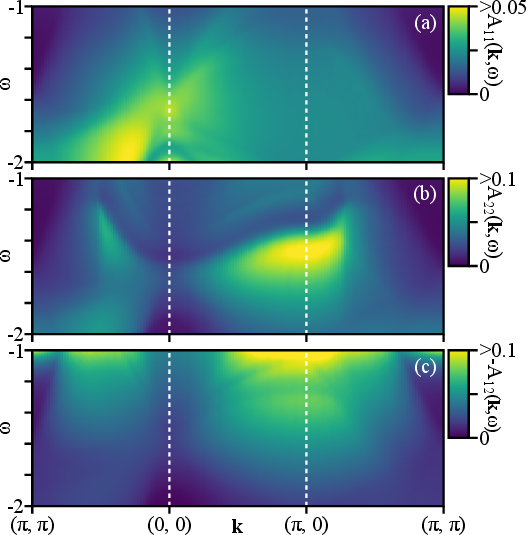}
\caption{{\bf Intensity maps of one-particle spectral function $\boldsymbol{A_{ab}(\vk, \omega)}$ by choosing a particular component of 
$\boldsymbol{{\rm Im}\Sigma_{ab}(\vk, \omega)}$.} (a) $a=b=1$, where the usual charge density-density correlations are considered, (b) $a=b=2$, which considers fluctuations associated with the local constraint, and (c) $a=1$ and $b=2$---the off-diagonal components. In (c), ${\rm Im}\Sigma_{12}(\vk, \omega)$ becomes positive in a negative $\omega$ and thus $A_{12}(\vk, \omega)$ becomes negative in $\omega<0$---the off-diagonal component alone is not physical. 
}
\label{ImSigab}
\end{figure}

We compute the spectral function for each component of Im$\Sigma_{ab}(\vk, \omega)$---we introduce the index in the spectral function as  $A_{ab}(\vk, \omega)$. Figure~\ref{ImSigab}(a)---the same as \fig{rAkw}d---shows the spectral function by considering only ${\rm Im}\Sigma_{11}(\vk, \omega)$. No plasmarons are present around $\omega = -1.5t$. If we consider only ${\rm Im}\Sigma_{22}(\vk, \omega)$, we then obtain rather broad plasmarons  in \fig{ImSigab}(b). While the intensity is high around $\vk=(\pi, 0)$, a dispersive feature of plasmarons is not so clear---compare \fig{ImSigab}(b) with \fig{replica}.  The off-diagonal component ${\rm Im}\Sigma_{12}(\vk, \omega)$ has the opposite sign to the diagonal components in $\omega<0$ as mentioned in Supplementary Note 2~A. This is the very reason why the plasmarons exhibit a rather sharp feature in Figs.~\ref{rAkw}a and \ref{replica} after summing up the off-diagonal components. For completeness, we present $A_{12}(\vk, \omega)$ in \fig{ImSigab}(c). 

\begin{figure}[th]
\centering
\includegraphics[width=8cm]{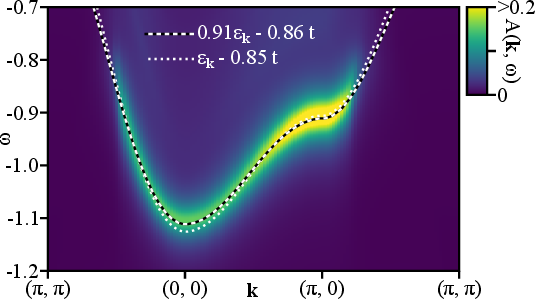}
\caption{{\bf Dispersion of plasmarons for $\boldsymbol{{\rm Sr_{0.9}La_{0.1}CuO_{2}}}$.} It follows $0.91  \varepsilon_{\vk} - 0.86t$ (dashed curve). 
The results can also be fitted equally by $\varepsilon_{\vk} -0.85t$ (dotted curve). 
}
\label{SLCO}
\end{figure}

\section*{Supplementary Note 3: Model parameters appropriate for $\boldsymbol{{\rm Sr_{0.9}La_{0.1}CuO_{2}}}$} 
While we have chosen the parameter set appropriate for LCCO in the main text, one may choose a different parameter set such as $t'/t=0.2$, $t_{z}/t=0.055$, $\alpha=4.1$, $V_{\rm c}/t=18.5$, and $\delta=0.10$, which was employed for infinite-layered electron-doped cuprate ${\rm Sr_{0.9}La_{0.1}CuO_{2}}$ \cite{hepting22}; the other parameters are the same as those for LCCO. We obtain the optical plasmon energy at $0.78t$. The resulting dispersion of plasmarons is shown in \fig{SLCO}, which is very similar to \fig{replica}. That is, the emergence of plasmarons does not depend on details of the parameter choice. Simply the energy of plasmons is expected to depend on the parameters, namely materials, and so is the energy of plasmarons---the renormalization factor of the plasmaron dispersion may also change.

\section*{Supplementary Note 4: Hole-doped case} 
\begin{figure}[bt]
\centering
\includegraphics[width=8cm]{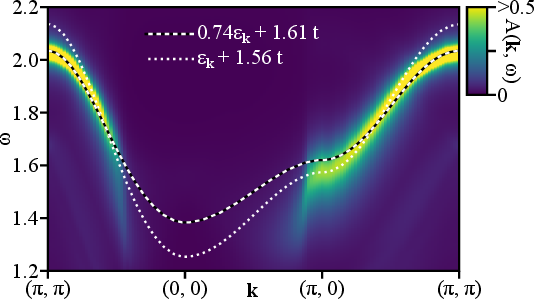}
\caption{{\bf Dispersion of plasmarons in hole-doped cuprates, especially $\boldsymbol{{\rm La_{2-x}Sr_{x}CuO_{4}}}$.} It follows $0.74 \varepsilon_{\vk} + 1.61t$ (dashed curve). 
The results may also be fitted by $\varepsilon_{\vk} + 1.56t$ (dotted curve). 
}
\label{hole}
\end{figure}

In the present theory of the $t$-$J$ model, the electron- and hole-doped cases are connected with each other via the particle-hole transformation \cite{yamase21a}: $\tilde{c}_{\vk \sigma} \rightarrow \tilde{c}_{\vk+\vQ \sigma}^{\dagger}$ and $\tilde{c}_{\vk \sigma}^{\dagger} \rightarrow \tilde{c}_{\vk+\vQ \sigma}$ with $\vQ=(\pi, \pi, 0)$. This transformation does not change the charge excitation spectrum  including plasmons, but $\vk$ and $\omega$ dependences of $\Sigma(\vk, \omega)$ and $A(\vk, \omega)$ change. Consequently, the incoherent band originating from plasmons is predicted in $\omega>0$ along the direction $(\pi,0)$-$(\pi, \pi)$-$(\pi/2, \pi/2)$ in the hole-doped case. Nonetheless no modifications occur in our conclusions---the incoherent band represents plasmarons, comes from coupling to the optical plasmon via fluctuations associated with the local constraint, namely from ${\rm Im}\Sigma_{22}(\vq, \omega)$, and is essentially the same as the so-called replica band discussed in weakly correlated electron systems \cite{aryasetiawan96,kheifets03,tediosi07,markiewicz07a,polini08,hwang08,bostwick10,brar10,walter11,guzzo11,dial12,lischner13,caruso15,caruso15a,lischner15,jang17,zliu21}. Since different materials can have different parameters, we have performed comprehensive studies for various choices of $t'$, $t_{z}$, $\alpha$, and $V_{\rm c}$, and confirmed that our conclusions are robust. Hence we can safely predict plasmarons also in hole-doped cuprates. Figure~\ref{hole} is the plasmaron dispersion obtained for $t'/t=-0.2$, $t_{z}/t=0.01$, $\alpha=3.5$, $V_{\rm c}=31$, $\delta=0.16$, and $t/2=0.35$ eV (the other parameters are the same as those for electron-doped cuprates; see Methods), which reproduce the plasmon spectrum in ${\rm La_{2-x}Sr_{x}CuO_{4}}$ with $x=0.16$ \cite{hepting22}, including the optical plasmon energy $\omega=1.2t$ \cite{uchida91}.

\section*{Supplementary Note 5: Possible plasmarons in RPA} 
The self-energy computed in the RPA [\eq{RPA}] has the same mathematical structure as ${\rm Im}\Sigma_{22}(\vq, \omega)$ [Eqs.~(\ref{ImSig}) and (\ref{ImDq0})] obtained in the $t$-$J$ model. It should be insightful to explore possible plasmarons in the RPA in the same context as the present work. 

We use the the same parameter set as the main text (see the last paragraph in Methods), but replace \eq{xik} by the usual tight-binding dispersion, 
\be
\varepsilon^{\rm RPA}_{\vk} = -2 t(\cos k_{x}+\cos k_{y})-4t'  \cos k_{x} \cos k_{y} -  2 t_{z} (\cos k_x-\cos k_y)^2 \cos k_{z} - \mu \,.
\label{xikRPA}
\ee 
With this dispersion we compute on-site charge fluctuations from particle-hole bubbles connected with the Coulomb interaction $V_{\vq}$, which can be summed algebraically and yield  
\be
\Pi^{\rm RPA} (\vq,  {\rm i} \nu_{n}) = \frac{\Pi_{0}(\vq,  {\rm i} \nu_{n})}{1+V_{\vq}\Pi_{0}(\vq,  {\rm i} \nu_{n})} \, ,
\ee
where $\Pi_{0}(\vq,  {\rm i} \nu_{n})$ is the Lindhard function \cite{mahan}. After the analytical continuation ${\rm i} \nu_{n} \rightarrow \nu+ {\rm i}  \Gamma_{\rm ch}$, we compute ${\rm Im}\Sigma^{\rm RPA}(\vq, \omega)$ given in \eq{RPA}.  

\begin{figure}[tb]
\centering
\includegraphics[width=8cm]{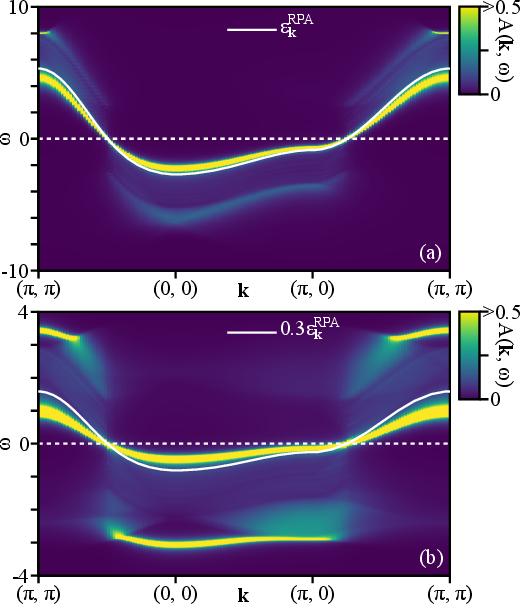}
\caption{{\bf Intensity maps of one-particle spectral function $\boldsymbol{A(\vk, \omega)}$ computed in RPA.}  $A(\vk, \omega)$ is shown along the direction $(\pi, \pi)$-$(0,0)$-$(\pi,0)$-$(\pi,\pi)$. (a) The white curve is the bare dispersion $\varepsilon^{\rm RPA}_{\vk}$ [\eq{xikRPA}]. (b) The same plot as (a), but a renormalized dispersion Eq.~(\ref{xikRPA2}) (white curve) is employed. 
}
\label{RPA-fig}
\end{figure}

Figure~\ref{RPA-fig}(a) is a map of $A(\vk, \omega)$. The strongest intensity corresponds to the quasiparticle dispersion. A comparison with the bare dispersion $\varepsilon^{\rm RPA}_{\vk}$ indicates a slight renormalization of the quasiparticle dispersion by the long-range Coulomb interaction. On top of that, two broad dispersive bands are realized  below and above the quasiparticle dispersion in a region $(\pi/2,\pi/2)$-$(0,0)$-$(\pi,0)$ and $(\pi,0)$-$(\pi,\pi)$-$(\pi/2,\pi/2)$, respectively---similar results were obtained in Ref.~\onlinecite{markiewicz07a}. These are damped plasmarons, which are realized with the long-range Coulomb interaction and disappear for the short-range Coulomb interaction. A rather sharp change of plasmarons from positive to negative energy around $(\pi,0)$ and $(\pi/2, \pi/2)$ is due to the term $n_{\mathrm{F}}( -\varepsilon_{\vk-\vq}^{\rm RPA}) +n_{\mathrm{B}}(\nu)$ in \eq{RPA} and occurs around the momentum where $\varepsilon^{\rm RPA}_{\vk}$ changes its sign, namely around the Fermi momentum. In the $t$-$J$ model, sharp plasmarons are realized only in the negative energy region as shown in the inset of Fig.~\ref{QP}. This is because the imaginary part of the self-energy is suppressed (enhanced)  by ${\rm Im}\Sigma_{12}(\vq, \omega)$ in the negative (positive) energy region (see Supplementary Note 2~A). Since this kind of mechanism is not present in the RPA or in a weakly correlated system,  plasmarons are realized in both positive and negative energy regions---they are in general damped and feature a faint structure as found previously  \cite{kheifets03,hwang08,markiewicz07a,lischner13,caruso15,caruso15a,lischner15}. 

This situation may change if the electron band width is small and thus the effect of the Coulomb interaction is relatively enhanced. To simulate this, we reduce the band width by $Z (<1)$ in \eq{xikRPA}: 
\be
\tilde{\varepsilon}^{\rm RPA}_{\vk} =Z \times \varepsilon^{\rm RPA}_{\vk} \, .
\label{xikRPA2}
\ee
Using this renormalized dispersion with $Z=0.3$, we obtain $A(\vk, \omega)$ shown in \fig{RPA-fig}(b). The energy scale of the quasiparticle dispersion is reduced approximately by $Z$.  Plasmarons are realized as features much sharper than \fig{RPA-fig}(a), similar to the case of the $t$-$J$ model shown in Figs.~\ref{rAkw}a and \ref{replica}, but in both positive and negative sides of $\omega$. Figure~\ref{RPA-fig} clearly suggests that plasmarons can be detected more easily in a narrow band system for weakly correlated materials where the RPA is expected to be reliable. In fact, the importance of the narrow band width to plasmarons was discussed in SrIrO$_{3}$ \cite{zliu21}. 

The optical plasmon energy is obtained as $2.52 t$ and $1.34 t$ in the present parameters for \eq{xikRPA} and \eq{xikRPA2}, respectively. While the plasmaron energy in \fig{RPA-fig}(a) is approximately given by a shift of the bare dispersion $\varepsilon^{\rm RPA}_{\vk}$ by the optical plasmon energy, such an approximation is not so good in \fig{RPA-fig}(b). As we have discussed for \fig{rAkw}e-h for $r=1$ and \fig{noqz0}, the optical plasmon is responsible for the peak structure of Im$\Sigma(\vk, \omega)$ and the peak energy is given by the optical plasmon energy plus the bare dispersion energy. The peak of Im$\Sigma(\vk, \omega)$ then leads to a dip structure in Re$\Sigma(\vk, \omega)$ slightly below the peak energy of  Im$\Sigma(\vk, \omega)$. It is the tail of this dip structure which determines the plasmaron energy given by $\omega - \varepsilon_{\vk} - {\rm Re}\Sigma(\vk, \omega)=0$. Hence in general the plasmaron energy deviates to some extent from the peak energy of Im$\Sigma(\vk, \omega)$. In many cases in the present work [see Figs~\ref{replica}, \ref{SLCO}, \ref{hole}, and \ref{RPA-fig}(a)], the plasmaron energy is well approximated by the peak energy of the Im$\Sigma(\vk, \omega)$, but this is not necessarily the case. 

\end{document}